\documentclass[a4paper]{jpconf}
\pdfoutput=1
\usepackage{graphicx}
\usepackage{braket}
\newcommand{\be}{\begin{equation}}
\newcommand{\ee}{\end{equation}}
\newcommand{\bea}{\begin{eqnarray}}
\newcommand{\eea}{\end{eqnarray}}

\usepackage{cite}
\usepackage[utf8]{inputenc} 
\usepackage{lmodern}
\begin{document}
\title{Features in single field slow-roll inflation}

\author{Alexander Gallego Cadavid}

\address{Instituto de F\'isica, Universidad de Antioquia, A.A.1226, Medell\'in, Colombia}

\ead{alexander.gallego@udea.edu.co}

\begin{abstract}
We compare the effects of local features (LF) and branch features (BF) of the inflaton potential on the spectrum of primordial perturbations. We show that LF affect the spectrum in a narrow range of scales while BF produce a step between large and small scales with respect to the featureless spectrum. We comment on the possibility of distinguishing between these two types of feature models from the analysis of the Cosmic Microwave Background (CMB) radiation data.

We also show that there exists a quantitative similarity between the primordial spectra predicted by two of the BF potentials considered. This could lead to a degeneracy of their predicted CMB temperature spectra which could make difficult to discriminate between the models from a CMB analysis. We comment on the possibility that the degeneracy can be broken when higher order terms in the perturbations are considered. In this sense non-Gaussianity may play an important role in discerning between different inflationary models which predict similar spectra.

\end{abstract}

\section{Introduction}
There is a significant set of observational data from experiments like the Wilkinson Microwave Anisotropy Probe (WMAP)~\cite{Peiris:2003ff,wmap,Hinshaw:2012aka,wmapfmr}, the Planck Satellite~\cite{Ade:2013sjv,Ade:2013zuv,Planck:2013jfk,Ade:2015xua,Ade:2015lrj}, the Sloan Digital Sky Survey (SDSS)~\cite{sdss}, and many other ground-based and sub-orbital experiments~\cite{nasa}. The surveys reveal that the observable universe is homogeneous and isotropic at large scales. 

The Cosmic Microwave Background (CMB) radiation is the relic radiation coming from the last scattering surface emitted when the universe was only about $380,000$ years old~\cite{wmapfmr,Ade:2013sjv,Ade:2013zuv,Ade:2015xua}. At this time, protons and electrons combined to form neutral light atoms (recombination) and photons started to travel freely through space (decoupling). Although this radiation is extremely isotropic, it is still possible to observe small fluctuations in the temperature map of the CMB radiation, with $\Delta T/T \sim 10^{-5}$~\cite{wmapfmr,Ade:2013sjv,Ade:2013zuv,Ade:2015xua}. Since at the recombination epoch the universe was very young, a detailed measurement and subsequent study of this radiation can give us valuable information about the physics of the early universe. 

The extraordinary advances in observable cosmology in the last decades give us evidence that the observed structures originated from seed fluctuations in the very early universe. Inflationary cosmology is one of the simplest frameworks able to explain the origin of these primordial fluctuations and to provide a good fit to the data~\cite{Planck:2013jfk,Ade:2015lrj,Guth:1980zm,Linde:1981mu,Starobinsky:1998mj,Martin:2013tda}. It is currently one of the most active research fields in describing the evolution of the very early universe. Cosmic inflation corresponds to an epoch of accelerated expansion of the universe. In the simplest models, inflation is driven by a single canonical scalar field, called the \emph{inflaton}, and slowly rolling a smooth potential~\cite{Planck:2013jfk,Ade:2015lrj}. During inflation the quantum fluctuations of the scalar field got stretched to cosmological scales which can later be seen as inhomogeneities and anisotropies in the Large Scale Structures (LSS) and the CMB radiation~\cite{Planck:2013jfk,Ade:2015lrj}.

The latest observations have led us to the Standard Model of cosmology, the so called $\Lambda$CDM model~\cite{wmapfmr,Ade:2013sjv,Ade:2013zuv,Ade:2015xua,Planck:2013jfk,Ade:2015lrj}. According to this model the universe is spatially flat, with \emph{nearly} Gaussian primordial perturbations with a spectrum usually taken to be a \emph{featureless} (smooth) primordial power spectrum (PPS), described by an \emph{almost} scale-invariant power-law~\cite{wmapfmr,Ade:2013sjv,Ade:2013zuv,Ade:2015xua}.

However, there exists several well-motivated theoretical inflationary models which predict a non-standard PPS with features~\cite{Leach:2001zf,Starobinsky:1992ts,Adams:2001vc,Chluba:2015bqa,Gariazzo:2015qea,Mooij:2015cxa,Appleby:2015bpw,Hunt:2015iua}. Indeed, we already have indications of primordial features in all releases of WMAP ~\cite{Peiris:2003ff,wmap,Hinshaw:2012aka,wmapfmr,Hazra:2016fkm} and Planck 2013 and 2015 temperature data~\cite{Ade:2015xua,Ade:2015lrj,Hazra:2016fkm}, and we have knowledge about the location and type of the features from reconstructions of the PPS \cite{Dorn:2014kga,Gariazzo:2014dla,Nicholson:2009zj,Hunt:2013bha}. 

Nowadays, primordial features is one of the most exciting extensions of the $\Lambda$CDM model~\cite{Chluba:2015bqa,Gariazzo:2015qea,Mooij:2015cxa,Appleby:2015bpw,Hunt:2015iua,Hazra:2016fkm,Gariazzo:2014dla,Arroja:2011yu,Romano:2014kla,Cadavid:2015iya,GallegoCadavid:2016wcz,GallegoCadavid:2015bsn,Benetti:2016tvm,Xu:2016kwz,Chen:2016vvw,DiValentino:2016ikp,Palma:2014hra,Martin:2014kja}. They can provide a wealth of information about the primordial universe, ranging from discrimination between inflationary models, distinguishing inflation from alternative scenarios, and even to new particle detection~\cite{Chluba:2015bqa,Gariazzo:2015qea,Mooij:2015cxa,Appleby:2015bpw,Hunt:2015iua,Hazra:2016fkm,Gariazzo:2014dla,Arroja:2011yu,Romano:2014kla,Cadavid:2015iya,GallegoCadavid:2016wcz,Xu:2016kwz}. Moreover it has been shown that features in the PPS can actually improve the fit to Planck CMB temperature and polarization data~\cite{Hazra:2016fkm,Adams:2001vc,Benetti:2016tvm,Motohashi:2015hpa,constraints1,constraints2,Hazra:2014jwa,Hunt:2013bha,Joy:2007na,Joy:2008qd,Mortonson:2009qv}

In this paper we are interested in an important class of models that predict primordial scale-dependent oscillatory features in the PPS. In particular, we study and compare the predictions on the PPS of two different kind of features of the inflaton potential. 

The paper is organized as follows. In Sec. \ref{m} we introduce the models and describe their different characteristics. In Sec. \ref{pscp} we calculate numerically the PPS for these models and discuss their different predictions. We conclude in Sec. \ref{c}.

\section{The inflationary models}\label{m}
We consider inflationary models with a single scalar field, minimally coupled to gravity, and standard kinetic term according to the action
\begin{equation}\label{action1}
  S = \int d^4x \sqrt{-g} \left[ \frac{1}{2} M^2_{Pl} R  - \frac{1}{2}  g^{\mu \nu} \partial_\mu \phi \partial_\nu \phi -V(\phi)
\right],
\end{equation}
where $g_{\mu \nu}$ is the Friedmann-Lemaitre-Robertson-Walker (FLRW) metric in a flat universe, $ M_{Pl} = (8\pi G)^{-1/2}$ the reduced Planck mass, $R$ the Ricci scalar, and $V$ is the potential energy of the inflaton $\phi$. The variation of the action with respect to the metric and the scalar field gives the Friedmann equation and the inflaton equation of motion respectively
\begin{equation}\label{ema}
  H^2 \equiv \left(\frac{\dot a}{a}\right)^2= \frac{1}{3 M^2_{Pl}}\left( \frac{1}{2} \dot \phi^2 + V(\phi) \right),
\end{equation}
\begin{equation}\label{emphi}
  \ddot \phi + 3H\dot \phi + \partial_{\phi}V = 0,
\end{equation}
where $H$ is the Hubble parameter, $a$ the scale factor, and we denote the derivatives with respect to time and scalar field with dots and $\partial_{\phi}$, respectively.

For the slow-roll parameters we use the following definitions
\bea \label{slowroll}
  \epsilon \equiv -\frac{\dot H}{H^2} \,\,\,\, , \,\,\,\, \eta \equiv \frac{\dot \epsilon}{\epsilon H}.
\eea
\begin{figure}
 \begin{minipage}{.45\textwidth}
  \includegraphics[scale=0.56]{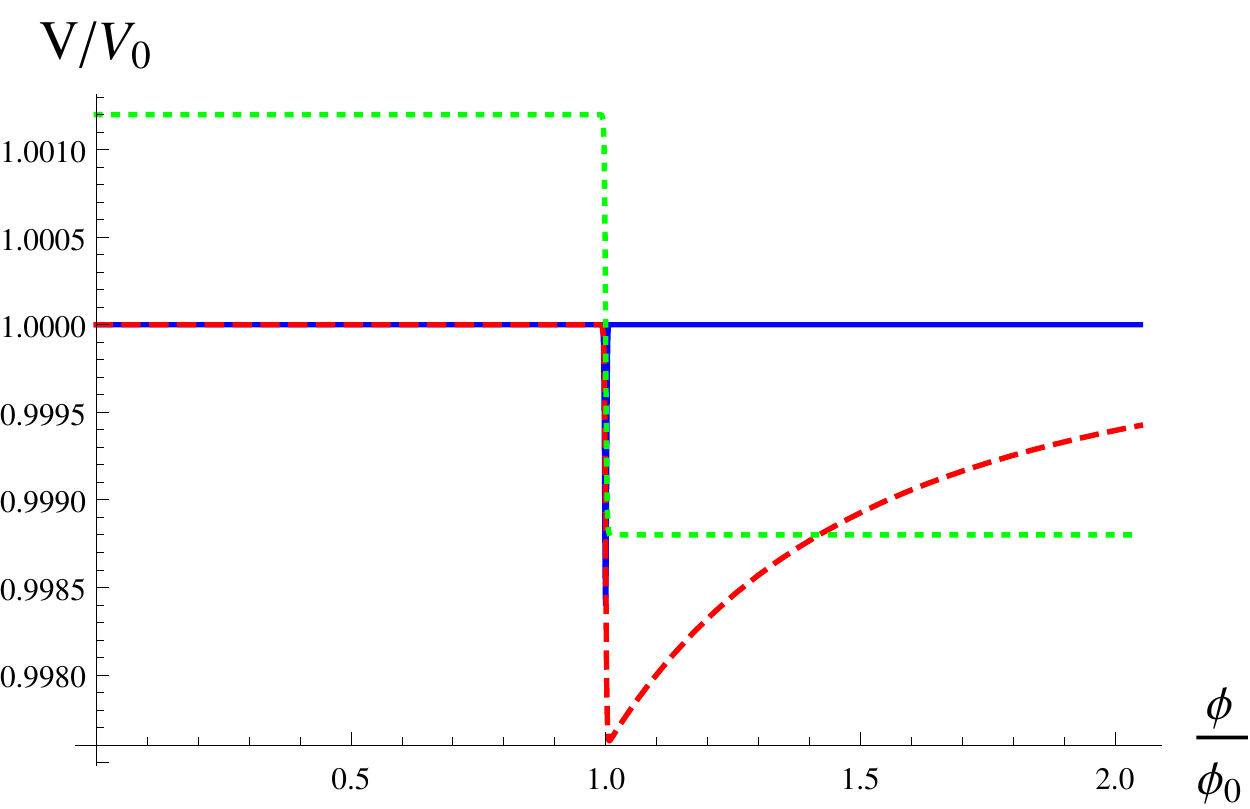}
  \end{minipage}
 \begin{minipage}{.45\textwidth}
  \includegraphics[scale=0.56]{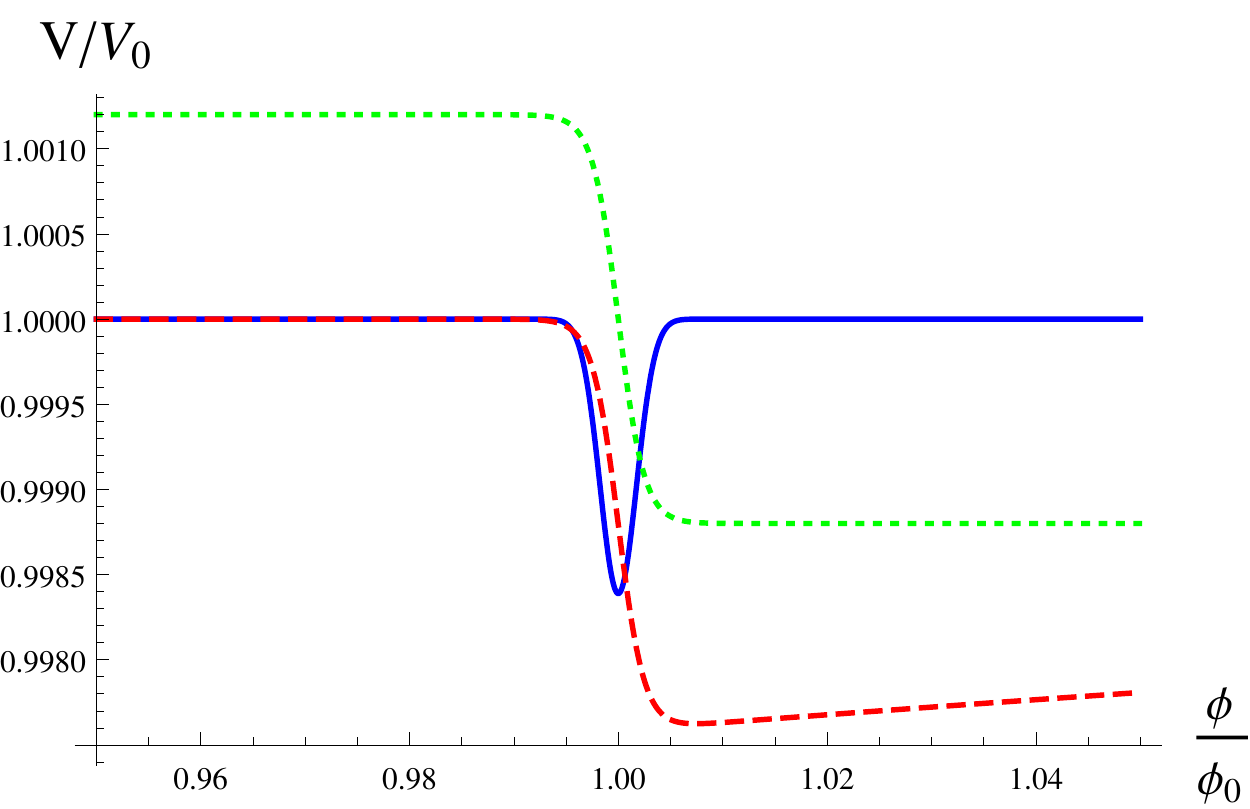}
 \end{minipage}
 \caption{The evolution of $V_i/V_0, (i=1,2,3)$ as a function of the field is plotted for the potentials in Eqs. (\ref{pot1}-\ref{pot3}). In all cases we use $\gamma=1.66 \times 10^{-11}, \sigma=0.04$, and we set the scale of the feature at $k_0=1.5 \times 10^{-3}$Mpc${}^{-1}$. For $V_1$, $V_2$, and $V_3$ we use $\lambda=-7\times10^{-12}$ (blue), $\lambda=-5.25 \times 10^{-12}$ (red-dashed), and $\lambda=-1.2 \times10^{-3}$ (green-dotted), respectively. On the left panel we can see that the potentials $V_2$ and $V_3$ are indeed different from each other for all values of the field. On the right panel we plot a zoom of the evolution around the feature value $\phi_0$.}
\label{Vplot}
\end{figure} 

\begin{figure}
   \begin{minipage}{.45\textwidth}
  \includegraphics[scale=0.56]{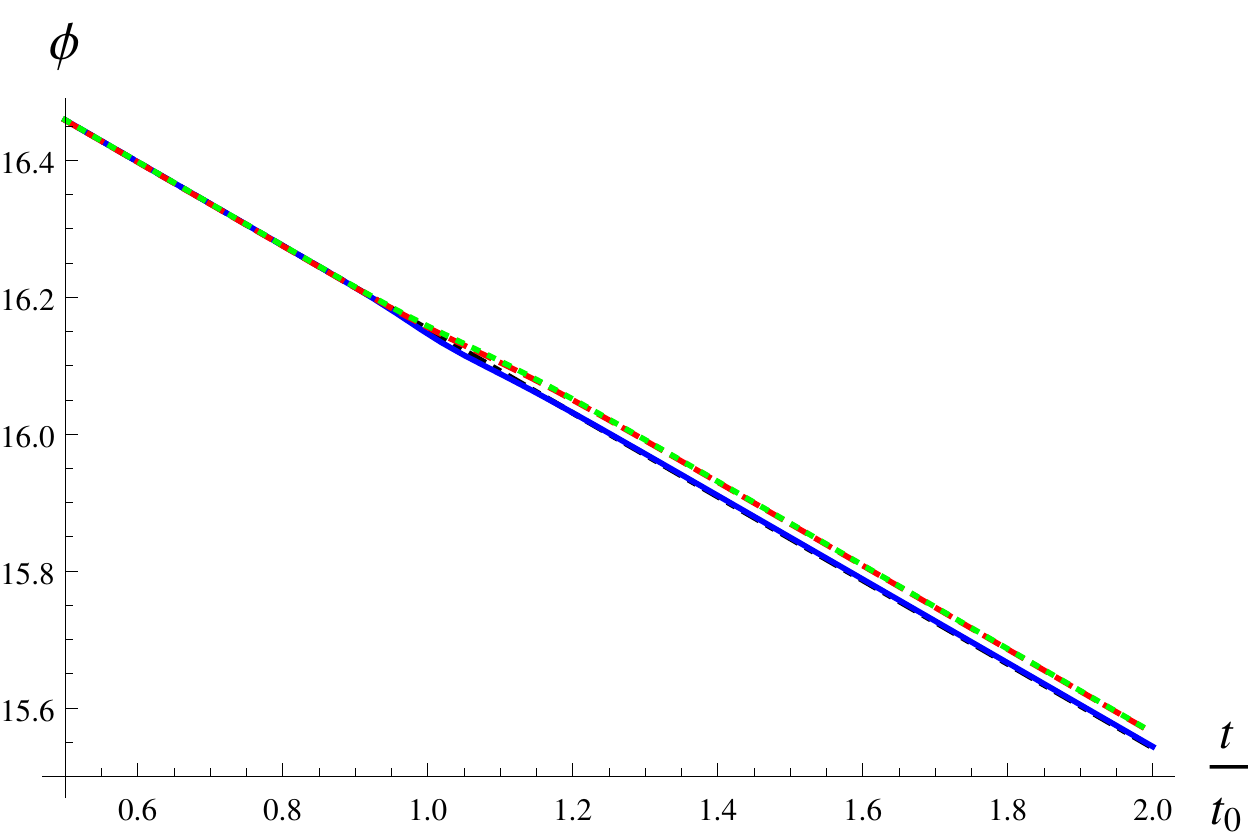}
  \end{minipage}
 \begin{minipage}{.45\textwidth}
  \includegraphics[scale=0.56]{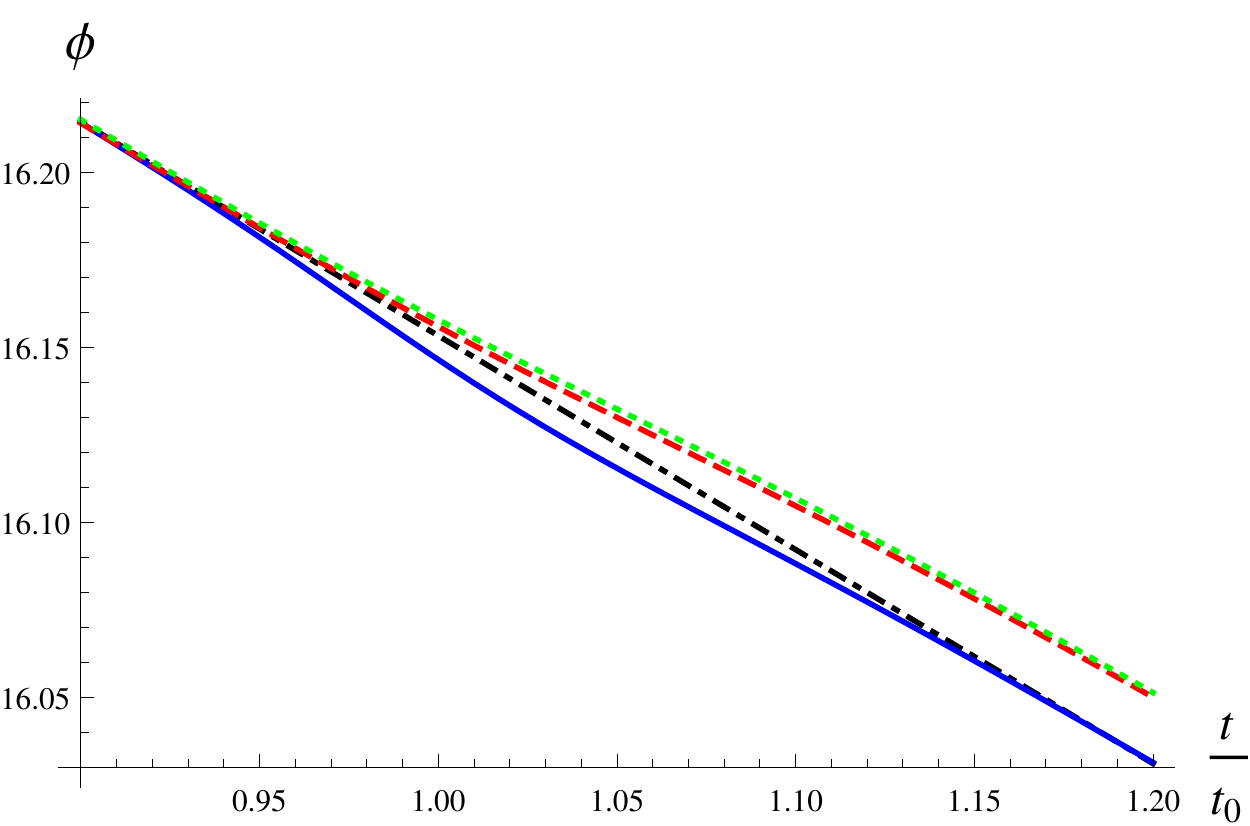}
 \end{minipage}
 \caption{The numerically computed inflaton evolution is plotted for the potentials in Eqs. (\ref{pot1}-\ref{pot3}). In all cases we use $\gamma=1.66 \times 10^{-11}, \sigma=0.04$, and we set the scale of the feature at $k_0=1.5 \times 10^{-3}$Mpc${}^{-1}$. For $V_1$, $V_2$, and $V_3$ we use $\lambda=-7\times10^{-12}$ (blue), $\lambda=-5.25 \times 10^{-12}$ (red-dashed), and $\lambda=-1.2 \times10^{-3}$ (green-dotted), respectively. The dashed-dotted black lines correspond to the featureless case. On the right panel we plot a zoom of the evolution around $\phi_0$.}
\label{phiplot}
\end{figure} 
 
\begin{figure}
 \begin{minipage}{.45\textwidth}
  \includegraphics[scale=0.56]{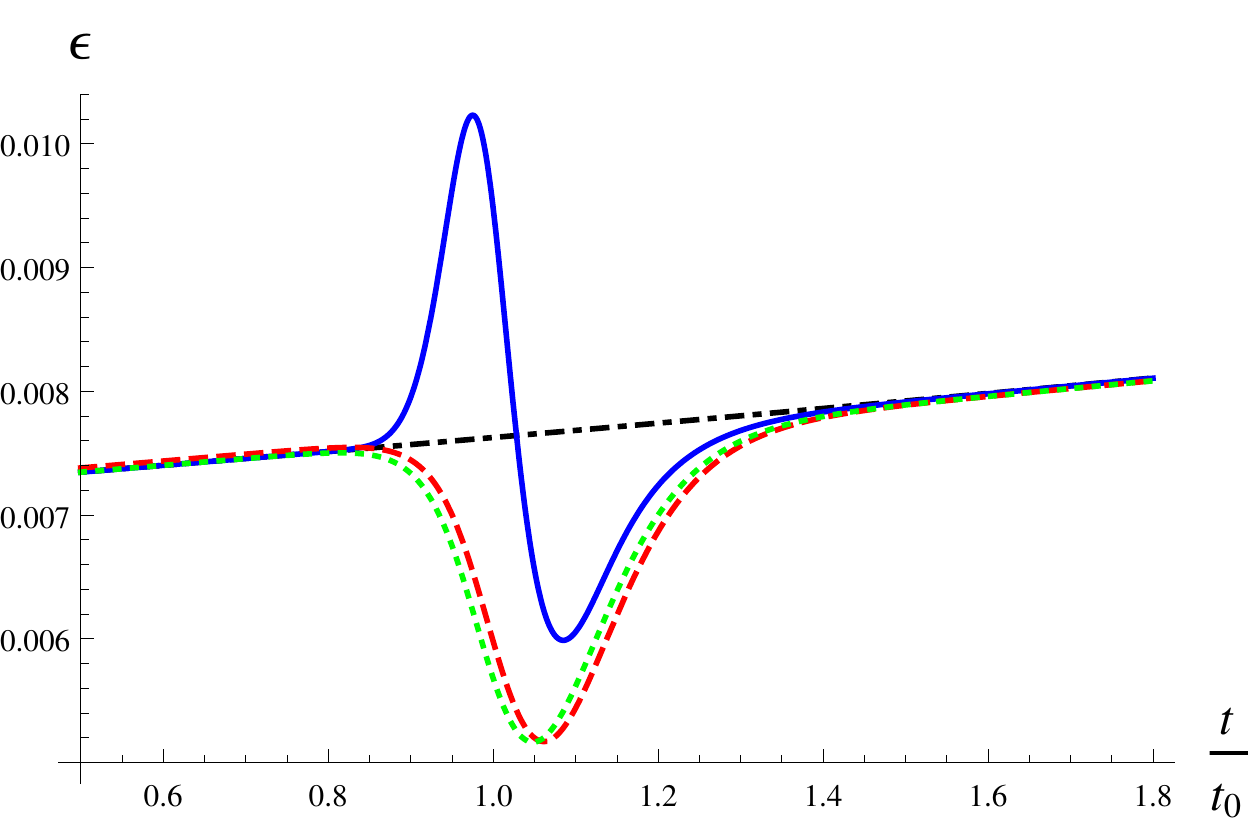}
  \end{minipage}
 \begin{minipage}{.45\textwidth}
  \includegraphics[scale=0.56]{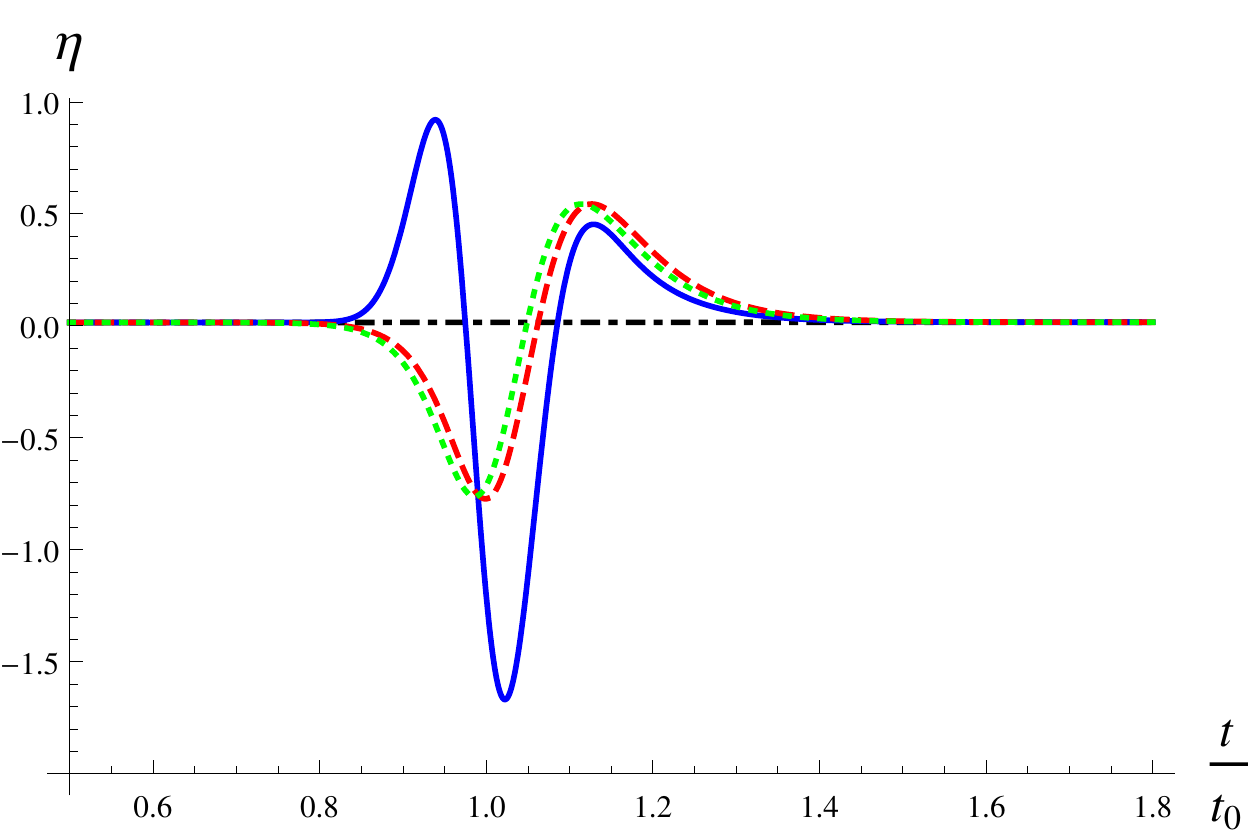}
 \end{minipage}
 \caption{The numerically computed slow-roll parameters are plotted for the potentials in Eqs. (\ref{pot1}-\ref{pot3}). In all cases we use $\gamma=1.66 \times 10^{-11}, \sigma=0.04$, and we set the scale of the feature at $k_0=1.5 \times 10^{-3}$Mpc${}^{-1}$. For $V_1$, $V_2$, and $V_3$ we use $\lambda=-7\times10^{-12}$ (blue), $\lambda=-5.25 \times 10^{-12}$ (red-dashed), and $\lambda=-1.2 \times10^{-3}$ (green-dotted), respectively. The dashed-dotted black lines correspond to the featureless case.}
\label{srplot}
\end{figure}
To study the effects of features on the spectrum of primordial perturbations we consider three different inflationary models defined by the potentials \cite{Cadavid:2015iya,Adams:2001vc,a1,a2}
  \bea
  V_1(\phi)&=& V_{0}(\phi) + V_{LF}(\phi) \, ,  \label{pot1}\\
  V_2(\phi)&=& V_{0}(\phi) + V_{BF2}(\phi) \, ,  \label{pot2}\\
  V_3(\phi)&=& V_{0}(\phi) V_{BF3}(\phi) \, , 	  \label{pot3}
  \eea
where $V_{0}$ is the chaotic inflaton~\cite{Ade:2015lrj,Planck:2013jfk} potential that we use as a toy model to show qualitatively the general type of effects produced by the features 
\bea
V_{0}(\phi)= \gamma \phi^2 \, ,
\eea
and
\bea\label{F}
V_{LF}(\phi)  &=& \lambda e^{-( \frac{\phi-\phi_0}{\sigma})^{2}} \, , \\
V_{BF2}(\phi) &=& \lambda \Biggl( 1+ \tanh  \Bigl(\frac{\phi-\phi_0}{\sigma} \Bigr) \Biggr) \, , \\
V_{BF3}(\phi) &=& 1+\lambda \tanh \Bigl(\frac{\phi-\phi_0}{\sigma} \Bigr)  \, .
\eea
In general the parameters $\lambda$ and $\sigma$ are related to the amplitude and width of the feature respectively~\cite{Arroja:2011yu,Cadavid:2015iya,Adams:2001vc,Romano:2014kla,a1,a2}. While the parameter $\phi_0 \equiv \phi(t_0)$, where $t_0$ is the feature time, is related to the scale of the feature $k_0 \equiv -1/\tau_0$, where $\tau$ is the conformal time $d\tau \equiv dt/a$~\cite{Arroja:2011yu,Cadavid:2015iya,Adams:2001vc,Romano:2014kla,a1,a2}. From now on we adopt a system of units in which $c=\hbar=M_{Pl}=1$.

The modifications of the slow-roll potential $V_1$ correspond to local features (LF), first studied in Ref.~\cite{Cadavid:2015iya}, while the ones in $V_2$ and $V_3$ are called branch features (BF) \cite{Arroja:2011yu,Cadavid:2015iya,Adams:2001vc,Romano:2014kla,Starobinsky:1992ts,Starobinsky:1998mj,Hazra:2014goa,a1,a2}. The BF differ from the LF because their definitions involve step functions or their smoothed versions, which divide the potential in separate branches~\cite{Cadavid:2015iya}. For these models the potential is not only modified around the feature $\phi_0$, but at any other field value after or before $\phi_0$~\cite{Cadavid:2015iya}, as shown in figure \ref{Vplot}. There it can be seen that the LF potential deviates from the featureless one only in a limited range of the field values around $\phi_0$ while the BF potentials differ from the featureless case: before the feature for potential $V_2$ and before and after the feature for $V_3$. In figure \ref{phiplot} we show the effects of the features on the inflaton evolution with respect to time. We see again that the BF diverge notoriously from the featureless behavior while the deviation of the LF is only around the feature time $t_0$. In figure \ref{srplot} we plot the effects of the features on the slow-roll parameters with respect to time.

Another interesting characteristic of potentials $V_2$ and $V_3$ is that, with the parameters chosen, they lead to a similar evolution of the inflaton and slow-roll parameters, as can be seen in figures \ref{phiplot} and \ref{srplot}. The importance of this behavior is that it could lead to a degeneracy of the primordial or CMB spectra which would make harder to distinguish between these different models.

\section{Primordial spectrum of curvature perturbations}\label{pscp}
The study of primordial scalar perturbations is attained by expanding perturbatively the action with respect to the background FLRW solution \cite{m}. In the comoving gauge, where the scalar field fluctuations vanish, $\delta \phi = 0$, the second order action for the scalar perturbations is~\cite{m}
\bea
\label{s2}
 S_2&=& \int dt d^3x\left[a^3 \epsilon \dot\mathcal{R}_{c}^2-a\epsilon(\partial \mathcal{R}_{c})^2 \right] \,,
\eea
where $\mathcal{R}_c$ is the curvature perturbation on comoving slices. The Euler-Lagrange equations for this action give
\begin{equation}
 \frac{\partial}{\partial t}\left(a^3\epsilon \frac{\partial \mathcal{R}_{c}}{\partial t}\right)- a\epsilon\delta^{ij} \frac{\partial^2 \mathcal{R}_{c}}{\partial x^i\partial x^j}=0.
\end{equation}
Then taking the Fourier transform we obtain the equation of motion for primordial curvature perturbation modes
\begin{equation}\label{cpe}
  \mathcal{R}_{c}''(k) + 2 \frac{z'}{z} \mathcal{R}_{c}'(k) + k^2 \mathcal{R}_{c}(k) = 0,
\end{equation}
where $z\equiv a\sqrt{2 \epsilon}$, $k$ is the comoving wave number, and primes denote derivatives with respect to conformal time. 

The Gaussianity of temperature fluctuations in the CMB implies that the statistical properties of the temperature field can be completely characterized by its two-point correlation function, i.e. its temperature spectrum~\cite{Ade:2015lrj,Planck:2013jfk}. In the case of curvature perturbations its two-point correlation function is given by~\cite{Ade:2015lrj,Planck:2013jfk}
\begin{equation}
 \Braket{ \hat \mathcal{R}_c(\vec{k}_1, t) \hat \mathcal{R}_c(\vec{k}_2, t) } \equiv (2\pi)^3 \frac{2\pi^2}{k^3} P_{\mathcal{R}_c}(k) \delta^{(3)}(\vec{k}_1+\vec{k}_2) \, ,
\end{equation}
where the power spectrum of primordial curvature perturbations is defined as 
\bea
P_{\mathcal{R}_{c}}(k) \equiv \frac{2k^3}{(2\pi)^2}|\mathcal{R}_{c}(k)|^2.
\eea
\begin{figure}
	\centering
	\includegraphics[scale=1.]{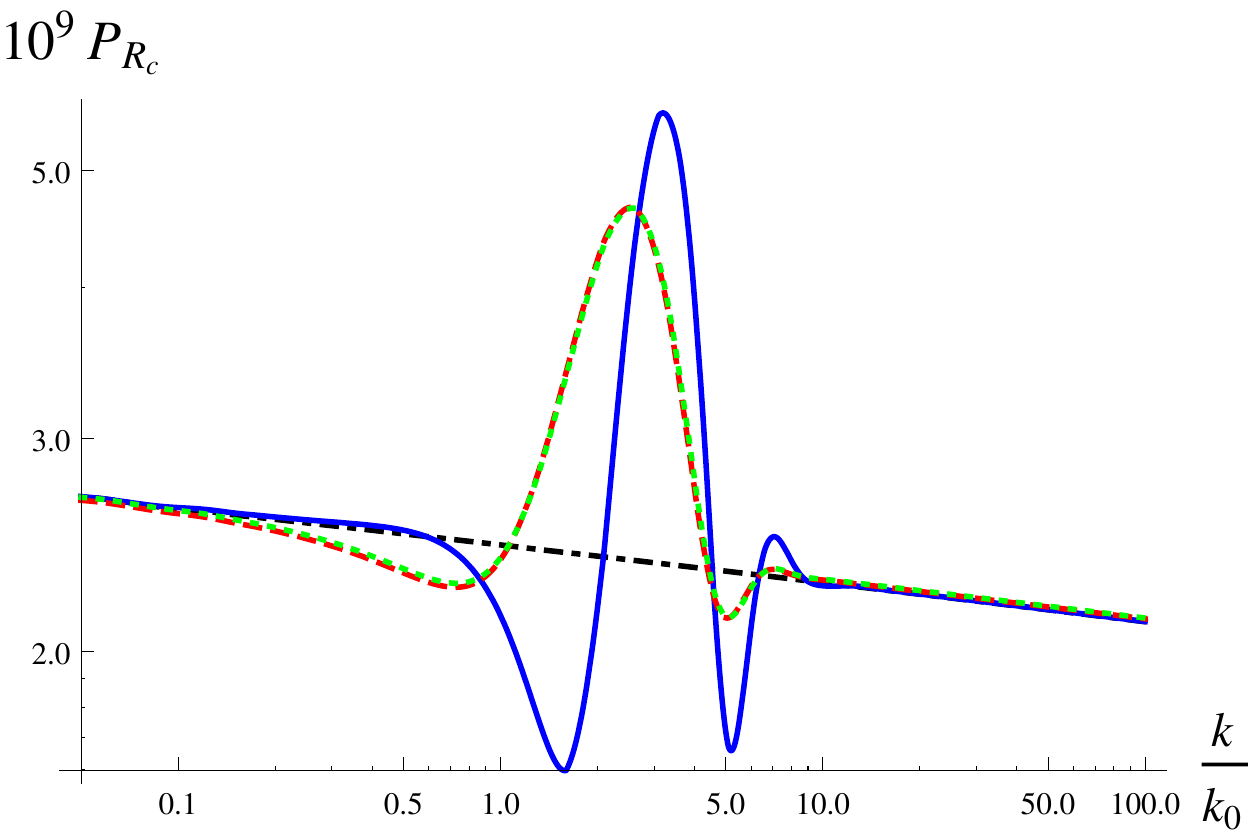}
	\caption{The numerically computed power spectrum of primordial curvature perturbations $P_{\mathcal{R}_{c}}$ is plotted for the potentials in Eqs. (\ref{pot1}-\ref{pot3}). In all cases we use $\gamma=1.66 \times 10^{-11}, \sigma=0.04$, and we set the scale of the feature at $k_0=1.5 \times 10^{-3}$Mpc${}^{-1}$. For $V_1$, $V_2$, and $V_3$ we use $\lambda=-7\times10^{-12}$ (blue), $\lambda=-5.25 \times 10^{-12}$ (red-dashed), and $\lambda=-1.2 \times10^{-3}$ (green-dotted), respectively. The dashed-dotted black lines correspond to the featureless spectrum.}
	\label{Pplot}
\end{figure}

\begin{figure}
 \begin{minipage}{.5\textwidth}
  \includegraphics[scale=0.62]{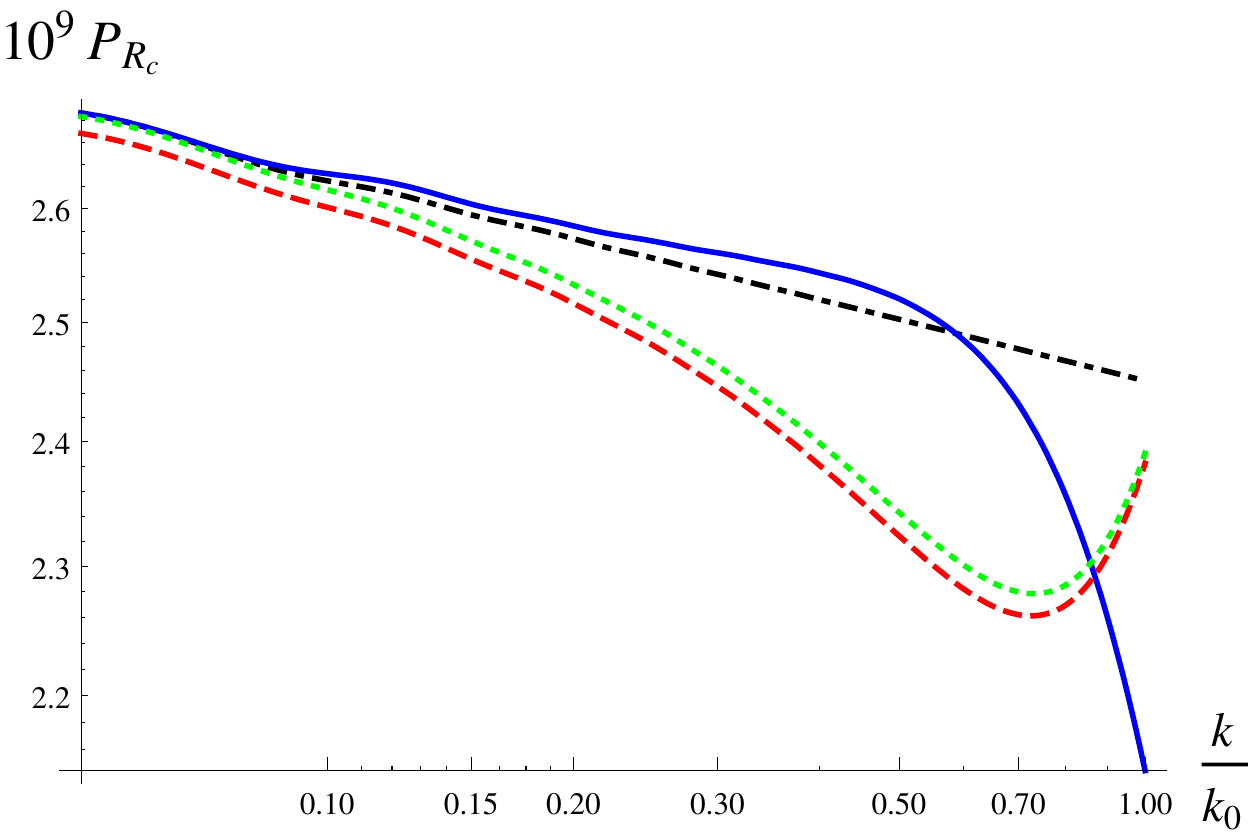}
  \end{minipage}
 \begin{minipage}{.5\textwidth}
  \includegraphics[scale=0.62]{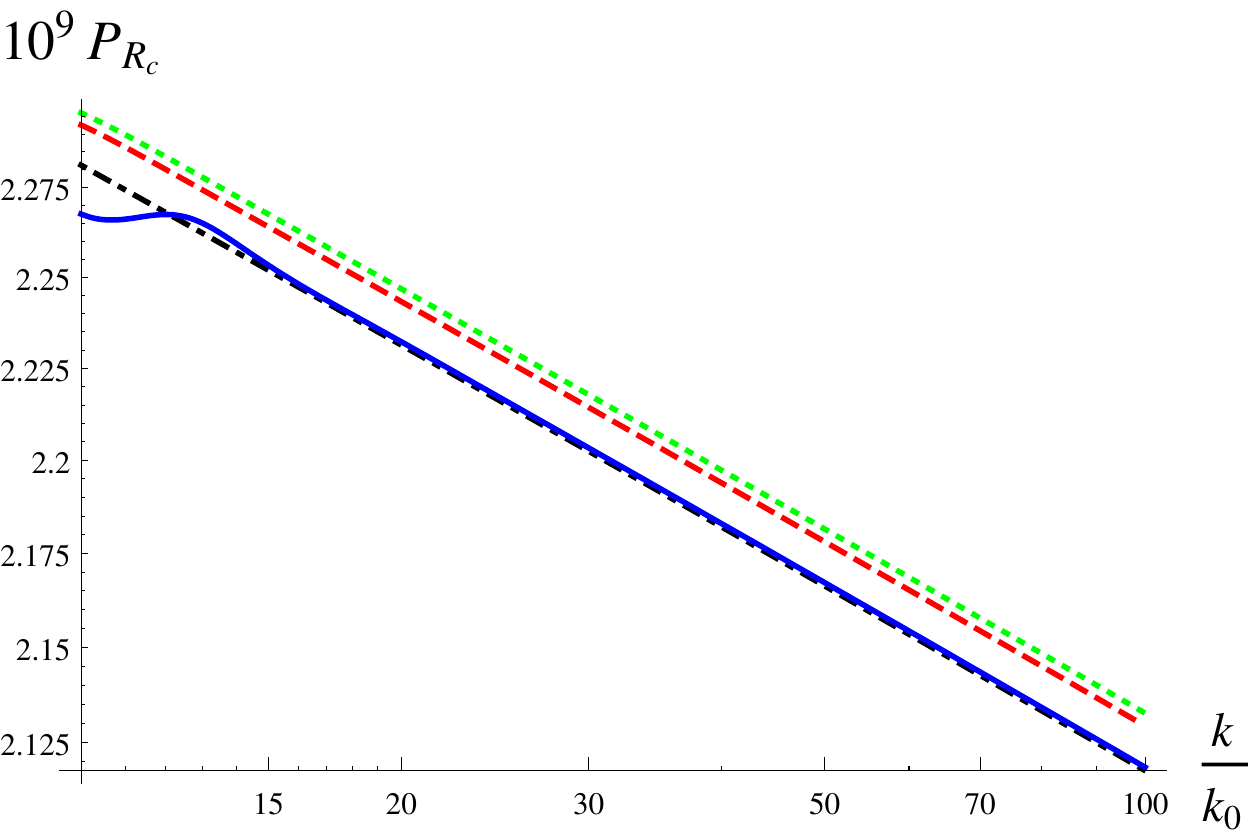}
 \end{minipage}
 \caption{The numerically computed power spectrum of primordial curvature perturbations $P_{\mathcal{R}_{c}}$ is plotted for the potentials in Eqs. (\ref{pot1}-\ref{pot3}). In all cases we use $\gamma=1.66 \times 10^{-11}, \sigma=0.04$, and we set the scale of the feature at $k_0=1.5 \times 10^{-3}$ Mpc${}^{-1}$. For $V_1$, $V_2$, and $V_3$ we use $\lambda=-7\times10^{-12}$ (blue), $\lambda=-5.25 \times 10^{-12}$ (red-dashed), and $\lambda=-1.2 \times10^{-3}$ (green-dotted), respectively. The dashed-dotted black lines correspond to the featureless spectrum. On the left (right) panel we make a zoom of the spectrum for large (small) scales.}
\label{Pplotzoom}
\end{figure}

We solve Eq.~(\ref{cpe}) numerically to obtain the spectrum of primordial curvature perturbations for the LF and BF. The different effects on the primordial spectrum are shown in figures \ref{Pplot} and \ref{Pplotzoom}. In figure \ref{Pplot} it can be seen that LF and BF produce similar oscillations of the spectrum around $k_0$~\cite{Cadavid:2015iya,Adams:2001vc,Arroja:2011yu,Romano:2014kla,Starobinsky:1992ts,Starobinsky:1998mj,Hazra:2014goa,a1,a2}. Nonetheless in figure \ref{Pplotzoom} we can see that BF have steps at large and small scales, i.e. differences between the feature and featureless spectra at large and small scales, which are absent for the LF~\cite{Cadavid:2015iya}. This indicates that the BF effects on the primordial perturbations are not only visible around the scale $k_0$ leaving the horizon around the feature time $\tau_0$, but for any other scale leaving the horizon when $\phi$ has a value within the feature branch~\cite{Cadavid:2015iya,Adams:2001vc,Arroja:2011yu,Romano:2014kla,Starobinsky:1992ts,Starobinsky:1998mj,Hazra:2014goa,a1,a2}. On the contrary LF only affect the perturbation modes which leave the horizon around $\tau_0$, and consequently the spectrum does not show a step, but a local dumped oscillation and then it approaches the featureless spectrum for sufficiently smaller and larger scales~\cite{Cadavid:2015iya}. This is crucial because it could allow to model local features of the temperature spectrum without affecting other scales~\cite{GallegoCadavid:2016wcz}. Moreover we could distinguish between LF and BF by the predictions of their effects on the temperature spectrum at large and small scales, where they are quantitatively different.

For the parameters $\lambda, \sigma$ and $k_0$ chosen, another important characteristic of the spectra obtained from $V_2$ and $V_3$ is that their behavior is very similar. In fact, from figure \ref{Pplot} we can see that the difference between the two spectra is quantitatively small. Only when we look at higher or smaller scales we can see a difference between them. If this difference is too small both primordial spectra could predict an indistinguishable CMB temperature spectrum which would make almost impossible to distinguish between the two different models.

It is also possible that this degeneracy could be easily broken at higher order terms, for instance by the three-point correlation function (or \emph{bispectrum})~\cite{Ade:2013ydc,Ade:2015ava}. Implying the importance of studying higher order correlation functions to gain more insight about the physics of the very early universe.

\section{Conclusions}\label{c}
Future CMB and LSS surveys will offer an impressive opportunity for the discovery of primordial features over the next decades. They would give us profound insight into the physics of the primordial universe such as probing the nature of inflationary models, distinguishing between inflation and alternative scenarios, and even discovering new particles.

In this paper we have studied the effects of LF and BF of the inflaton potential on the spectrum of primordial curvature perturbations. 

We have seen that LF affect the spectrum in a narrow range of scales, as opposed to BF which produce a step with respect to the featureless spectrum at large and small scales. It would be interesting to study if the differences between the LF and BF spectra at large and small scales can actually be quantified in a CMB spectrum analysis.

Another important result is the similarity between the primordial spectra predicted by the two BF potentials considered. This could lead to a degeneracy of the CMB temperature spectrum which could make even harder the discrimination between the models from a CMB analysis. It is possible though that the degeneracy can be broken when higher order perturbation terms are considered, such as the bispectrum. In this sense non-Gaussianity may play an important role in discerning between different inflationary models which predict similar spectra. It is left for a future work the study of such a degeneracy breaking for the models considered in this paper. 

As a final remark, it would be interesting to study phenomenologically a singled potential containing both LF and BF and study its predictions on the primordial and CMB spectra.

\ack
This work was supported by the Colombian Department of Science, Technology, and Innovation COLCIENCIAS research Grant No. 617-2013. A.G.C. acknowledges the partial support from the International Center for Relativistic Astrophysics Network ICRANet during his stay in Italy and valuable comments from Clément Stahl.
\section*{References}
\bibliography{Bibliography}
\bibliographystyle{iopart-num}

\end{document}